\documentstyle [11pt,epsf] {article}

\catcode`@=12
\newcommand{\be}{\begin{equation}}
\newcommand{\ee}{\end{equation}}
\newcommand{\ber}{\begin{eqnarray}}
\newcommand{\eer}{\end{eqnarray}}
\newcommand{\bers}{\begin{eqnarray*}}
\newcommand{\eers}{\end{eqnarray*}}

\begin{document}
\vspace{0.5in}
\oddsidemargin -.375in  
\newcount\sectionnumber 
\sectionnumber=0 
\def\epjc#1#2#3{{ Eur.\ Phys.\ J.}\ {\bf C#1}, #3, (#2)} 
\def\ijmp#1#2#3{{ Int.\ J.\ Mod.\ Phys.} {\bf A#1}, #3 (#2)}
\def\mpla#1#2#3{{Mod.\ Phys.\ Lett.} {\bf A#1}, #3 (#2)}
\def\nci#1#2#3{{ Nuovo Cimento} {\bf #1}, #3 (#2)}
\def\npb#1#2#3{{ Nucl.\ Phys.} {\bf B#1}, #3 (#2)}
\def\plb#1#2#3{{ Phys.\ Lett.} {\bf #1B}, #3 (#2)}
\def\prd#1#2#3{{ Phys.\ Rev.} {\bf D#1}, #3 (#2)}
\def\newprd#1#2#3{{ Phys.\ Rev.} {\bf D#1}, #3 (#2)}
\def\prl#1#2#3{{ Phys.\ Rev.\ Lett.} {\bf #1}, #3 (#2)}
\def\newprl#1#2#3{{ Phys.\ Rev.\ Lett.} {\bf #1}, #3 (#2)}
\def\zpc#1#2#3{{ Zeit.\ Phys.} {\bf C#1}, #3 (#2)}

\def\be{\begin{equation}} 
\def\ee{\end{equation}}
\thispagestyle{empty}
\vspace {.5in}
\begin{center}
{\Large \bf{Understanding the nature of $D_s(2317)$ and
$D_s(2460)$ through nonleptonic 
 B Decays\\}}
\vspace{.5in}
{\rm { Alakabha Datta$^1$ and  
 P. J. O'Donnell$^1$ } \\}

\vskip .3in

{\it $^1$Department of Physics  \\}
{\it  University of Toronto, \\}
{\it Toronto, Ontario, Canada M5S 1A7}\\

\vskip .5in
\end{center}  
\begin{abstract}
  
  We consider the nonleptonic $B$ decays $ B \to D^{(*)} D_s(2317)$
  and $ B \to D^{(*)} D_s(2460)$, involving the newly discovered
  $D_s(2317)$ and the $D_s(2460)$ states.  We find that experiments
  indicate disagreement with model calculations of their properties
  and/or breakdown of the factorization assumption for these decays .
  We point out that decays involving $B_s$ mesons where the $D_s$
  resonances can be produced via the weak decay of the $b$ quark can
  provide further information about the nature of these newly
  discovered states.  We also propose a model to calculate the two
  body nonleptonic decays $ B \to D^{(*)} D_s(2317)(D_s(2460))$, if
  the $D_s(2317)$ and $D_s(2460)$ are interpreted as $DK$ and $D^*K$
  molecules.

\end {abstract}

\newpage
\baselineskip 24pt

%
%
%

\section{Introduction}
There has been recent observations of an unexpectedly light narrow
resonance in $D^+_s\pi^0$ with a mass of $2317 MeV/c^2$ by the BaBar
collaboration \cite{babarDs}, together with another second narrow
resonance in $D_s\pi^0\gamma$ with a mass $2460 MeV/c^2$
\cite{cleoDs}.

The smaller than expected masses and narrow widths of these states
have led, among other explanations \cite{jackson}, to a multi-quark
anti-quark or a $D K$ molecule interpretation of these states
\cite{barneslipkin}, or to an interpretation as p-wave states where
the light degrees of freedom are in an angular momentum state $j_q =
{1 \over 2}$ \cite{bardeenhill}, or even some combination of these
\cite{sandip}.  There are also conflicting lattice interpretations of
these states \cite{latticeDs}.  The mass difference between the
$D_s(2317)$ and the well established lightest charm-strange meson,
$D_s$, is $\Delta M = 350 MeV/c^2$. This is less than the kaon mass,
thus kinematically forbidding the decay $D_s(2317) \rightarrow
D_{u,d}+K$.  The possible resonance at $2460 MeV/c^2 $ also has such a
mass difference when taken with the lighter $D^*$ state.  The
interpretation of these states as bound $D^{(*)} K$ molecules just
below the $D^{(*)}K$ threshold is particularly interesting in the
light of the recent discovery of a narrow resonance in the decay
$J/\psi \rightarrow \gamma p \bar{p}$ \cite{bes} which has been
interpreted as a zero baryon number, ``deuteron-like singlet $ {}^1S_0
$'' bound state of $p$ and $\bar{p}$ \cite{datta-baryonium}.

In the heavy quark theory, the ground state heavy meson involving a
heavy and a light quark has the light degrees of freedom in a
spin-parity state $j^P_q={1\over 2}^-$, corresponding to the usual
pseudoscalar-vector meson doublet with $J^P=(0^-,1^-)$.  The first
excited states involves a p-wave excitation, in which the light
degrees of freedom have $j^P_q={1 \over 2}^+$ or ${3\over2}^+$.  This
leads to two heavy doublets, the first giving $J^P=(0^+,1^+)$ and the
latter a heavy doublet with $J^P=(1^+,2^+)$.  Heavy quark symmetry
rules out any pseudoscalar coupling of this doublet to the ground
state at lowest order in the chiral expansion \cite{Fluke} and so
these states are expected to be narrow. Recent Belle analysis of $ B
^- \to D^{(+ *)} \pi^+ \pi^-$ decays \cite{belletrabelsi} indicate the
presence of the $1^+$ state in this multiplet at a mass of $M_{D_1^0}=
(2421.4 \pm 2.0 \pm 0.4 \pm 0.8) MeV/c^2 $ with a width of
$\Gamma_{D_1^0}= (23.7 \pm 2.7 \pm 0.2 \pm 4.0)$ MeV .  The other
state in the doublet ($2^+$) is also found with a mass of $M_{D_2^0}=
(2461.6 \pm 2.1 \pm 0.5 \pm 3.3) MeV/c^2$ with a width of
$\Gamma_{D_1^0}= (45.6 \pm 4.4 \pm 6.5 \pm 1.6)$ MeV .  In the $D_s$
system the counterpart states to these are naively expected to be a
$100$ MeV heavier because of the strange quark mass and so these
states can probably be identified with $D_{s1}(2536)$ and
$D_{sJ}(2573)$ \cite{PDG}.  This is in line with the experimental
observations that in the ground state the $D_s$ mesons are about a 100
MeV heavier than their nonstange counterparts.

The other excited doublet has $J^P=(0^+,1^+)$.  These states are
expected to decay rapidly through s-wave pion emission in the
$D_{u,d}$ system and by kaon emission in the $D_s$ system and have
large widths \cite{nathanmark}. Observation of the $1^+$ state in the
$D$ system was reported by CLEO \cite{cleores} some time ago.  The
recent Belle analysis of $ B ^- \to D^{(+ *)} \pi^+ \pi^-$ decays
\cite{belletrabelsi} also find evidence for the states in this doublet
at $M_{D_0^{*0}}(0^+)= (2308 \pm 17 \pm 15 \pm 20) MeV/c^2$ with a
width of $\Gamma_{D_1^0}= (276 \pm 21 \pm 18 \pm 60)$ MeV .  The other
state in the doublet is also found with a mass of $M_{D_1^{*0}}(1^+)=
(2427 \pm 26 \pm 20 \pm 15) MeV/c^2$ with a width of $\Gamma_{D_1^0}=
(384^{+107}_{-75} \pm 24 \pm 70)$ MeV .  Note that these states are broad
as expected from theory. Naively then, we should expect the $D_s$
counterparts of these states at $M_{D_s}(0^+) \approx 2408$ and
$M_{D_s}(1^+) \approx 2527$. These numbers are consistent with quark
model estimates \cite{godfreyK} and we expect these states to be
broad.  The recently observed $D_s$ resonances have masses below these
expectations and are very narrow, decaying through isospin violating
transitions to $D_s^{(*)} \pi$ final states. This has generated
speculations that these states may not be p-wave excited states but
rather something exotic like $D^{(*)}K$ molecules.

While the spectroscopy of these newly discovered states can provide
clues to their structure, decays involving these states can yield
further clues to their exact nature. We first look at nonleptonic B
decays involving the p-wave $D_s$ resonant states which we will denote by
$D_{s0}$, corresponding to the p-wave, $j_q={1 \over 2}$, $0^+$ state,
and $D_{s1}^*$ corresponding to the p-wave, $j_q={1 \over 2}$, $1^+$
state. In $B$ factories that do not produce the $B_s$ mesons the $D_s$
p-wave states cannot be directly produced via the weak current
involving the $b$ quark but they can only be produced through the
$\overline{s} c$ current in the weak decay effective Hamiltonian. It
was suggested in Ref. \cite{datta-beta,colangelo1} that these
theoretically expected broad states may be discovered through the
three body decays $B \to D^{(*)} D^{(*)} K$ decays, where $D^{(*)}$
refer to $D$ or $D^*$, if these states are above the $D^{(*)}K$
threshold. These three body decays can also be used to measure both
$\sin{ 2\beta}$ and $\cos{ 2 \beta}$
\cite{charles,colangelo2,datta-beta}.  In hadron B factories the $D_s$
resonant states can be produced directly from the weak decay of the
$b$ quark in the $B_s$ meson.

In this work we concentrate on non leptonic decays of the type $B \to
D^{(*)}D_s(2317)$ and $B \to D^{(*)}D_s(2460)$, which are accessible at
current $B$ factories, and we also study nonleptonic decays of the
types $B_s \to D_s(2317) M$ and $B_s \to D_s(2460)M$, where $M$ is the
meson formed by the emitted $W$. These latter decays can be studied at
hadron $B$ factories. Our purpose here is to explore what additional
information about the structure and the properties of 
the new $D_s$ states can be
obtained from these nonleptonic decays.

\section{Nonleptonic Decay}
Let us first assume that we can identify the the newly discovered states
$D_s(2317)$ with $D_{s0}$ and $D_s(2460)$ with $D_{s1}^*$.
In the Standard Model (SM) 
the amplitudes for $B \to D^{(*)}D_{s0}(D_{s1}^*)$, 
are generated by the following effective 
Hamiltonian \cite{buras}:
\begin{eqnarray}
H_{eff}^q &=& {G_F \over \protect \sqrt{2}} 
   [V_{fb}V^*_{fq}(c_1O_{1f}^q + c_2 O_{2f}^q) -
     \sum_{i=3}^{10}(V_{ub}V^*_{uq} c_i^u
+V_{cb}V^*_{cq} c_i^c +V_{tb}V^*_{tq} c_i^t) O_i^q] +H.C.\;,
\end{eqnarray}
where the
superscript $u,\;c,\;t$ indicates the internal quark, $f$ can be $u$ or 
$c$ quark, $q$ can be either a $d$ or a $s$ quark depending on 
whether the decay is a $\Delta S = 0$
or $\Delta S = -1$ process.
The operators $O_i^q$ are defined as \cite{datta-dcp}
\begin{eqnarray}
O_{1f}^q &=& \bar q_\alpha \gamma_\mu Lf_\beta\bar
f_\beta\gamma^\mu Lb_\alpha\;,\;\;\;\;\;\;O_{2f}^q =\bar q
\gamma_\mu L f\bar
f\gamma^\mu L b\;,\nonumber\\
O_{3,5}^q &=&\bar q \gamma_\mu L b
\bar q' \gamma_\mu L(R) q'\;,\;\;\;\;\;\;\;O_{4,6}^q = \bar q_\alpha
\gamma_\mu Lb_\beta
\bar q'_\beta \gamma_\mu L(R) q'_\alpha\;,\\
O_{7,9}^q &=& {3\over 2}\bar q \gamma_\mu L b  e_{q'}\bar q'
\gamma^\mu R(L)q'\;,\;O_{8,10}^q = {3\over 2}\bar q_\alpha
\gamma_\mu L b_\beta
e_{q'}\bar q'_\beta \gamma_\mu R(L) q'_\alpha\;,\nonumber
\end{eqnarray}
where $R(L) = 1 \pm \gamma_5$, 
and $q'$ is summed over all flavors except t.  $O_{1f,2f}$ are the 
current-current operators that represent tree level processes. $O_{3-6}$ are the strong gluon induced
penguin operators, and operators 
$O_{7-10}$ are due to $\gamma$ and Z exchange (electroweak penguins),
and ``box'' diagrams at loop level. The values of the Wilson coefficients
can be found in Ref. \cite{buras}.

In the factorization assumption the amplitude for 
$B \to D^{(*)}D_{s0}(D_{s1}^*)$,  can now be written as
\be
M=M_1 +M_2\
\ee
where 
\ber
M_1 & = &\frac{G_F}{\protect \sqrt{2}} X_1
 <D_{s0}(D_{s1}^*)|\, \bar{s} \gamma_\mu(1-\gamma^5) \, c\, |\, 0>
               <D^{(*)} |\, \bar{c} \, \gamma^\mu (1-\gamma^5) \, b\, |\, B>
\nonumber\\
M_2 & = &\frac{G_F}{\protect \sqrt{2}} X_2
 <D_{s0}(D_{s1}^*)|\, \bar{s} (1+\gamma^5) \, c\, |\, 0>
               <D^{(*)} |\, \bar{c} \, (1-\gamma^5) \, b\, |\, B>  \\
\eer
where
\ber
        X_1 & = & V_c \left(\frac{c_1}{N_c} + c_2 \right) + 
                  \frac{B_3}{N_c} + B_4
                + \frac{B_9}{N_c} + B_{10} \nonumber\\
        X_2 & = & -2 \, \left(\frac{1}{N_c} B_5 + B_6 + \frac{1}{N_c} B_7 
                + B_8 \right)
\eer
We have defined
\be
        B_i = - \sum_{q=u,c,t} c_i^q V_q
\label{nonleptonic}
\ee
with
\be
        V_q = V_{qs}^{*} V_{qb}
\ee

In the above equations $N_c$ represents the number of colors.  To
simplify matters we neglect the small penguin contributions and so as
a first approximation we will neglect $M_2$.  The currents involving
the heavy $b$ and $c$ quarks, $J^{\mu}_{D} =<D|{\overline c}
\gamma^{\mu}(1-\gamma_5)b|B(p)> $ and $J^{\mu}_{D^{*}}
=<D^{*}(\epsilon_{1})|{\overline c} \gamma^{\mu}(1-\gamma_5)b|B(p)> $
can be expressed in terms of form factors \cite{BSW}.  In the heavy
quark limit the various form factors are related to a universal
Isgur-Wise function $\xi(v\cdot v_1)$ where $v$ and $v_1$ are the four
velocities of the $B$ and the $D^{(*)}$ mesons.  One can therefore
write,
\ber
J^{\mu}_{D} & = & \sqrt{m_B}\sqrt{m_D}\xi(v \cdot v_1)
\left[v^{\mu}+ v_1^{\mu} \right] \
\label{D}
\eer
and
\ber
J^{\mu}_{D^*} & = & \sqrt{m}\sqrt{m_1}\xi(v\cdot v_1)
\left[-i\varepsilon^{\mu\nu\alpha\beta}
\epsilon^{*}_{1\nu}v_{\alpha}v_{1\beta}
+v^{\mu}_{1}\epsilon^{*}_1\cdot v -\epsilon^{*\mu}_{1}(v\cdot v_1 +1) 
\right] \
\label{Dstar}
\eer

The matrix elements
$ <D_{s0}|\, \bar{s} \gamma_\mu(1-\gamma^5) \, c\, |\, 0>$
and $ <D_{s1}^*|\, \bar{s} \gamma_\mu(1-\gamma^5) \, c\, |\, 0>$ are
written in terms of the decay constants that are defined as
\ber
<D_{s0}(P)|\, \bar{s} \gamma_\mu(1-\gamma^5) \, c\, |\, 0>
& = &i f_{D_{s0}} P_{\mu}\nonumber\\
<D_{s1}^*(P, \varepsilon_2)|\, \bar{s} \gamma_\mu(1-\gamma^5) \, c\, |\, 0>
& = & M_{D_{s1}^*} f_{D_{s1}^*} \varepsilon^{*}_{2\mu}\
\eer

We can now define the following ratios
\ber
R_{D0}= \frac{BR[B \to D D_{s0}]}{BR[B \to D D_s]} \nonumber\\
R_{D^*0}= \frac{BR[B \to D^* D_{s0}]}{BR[B \to D^* D_s]} \nonumber\\
R_{D1}= \frac{BR[B \to D D_{s1}^*]}{BR[B \to D D_s^*]} \nonumber\\
R_{D^*1}= \frac{BR[B \to D^* D_{s1}^*]}{BR[B \to D^* D_s^*]} \
\label{ratios}
\eer
Let us focus on the ratio $R_{D0}$ which within factorization and the 
heavy quark limit can be 
written as
\ber
R_{D0} & = & |\frac{f_{D_{s0}}}{f_{D_s}}|^2\
\eer
where we have neglected phase space ( and other) effects that are
subleading in the heavy quark expansion.
Similarly we have
\ber
R_{D1} & = & |\frac{f_{D_{s1}^*}}{f_{D_s^*}}|^2\
\eer

Now in the heavy quark limit $ f_{D_{s0}} = f_{D_{s1}^*}$ and $
f_{D_{s}} = f_{D_{s}^*}$ and so one would predict $R_{D0}  \approx 
R_{D1}$. There have been various estimates of the decay constant
$f_{D_{s0}}$ in quark models \cite{decayconstantQM} and in QCD sum
rule calculations \cite{decayconstantQCDsumrule}; these typically find
the p-wave , $j_q= {1 \over 2}$ states to have the similar decay
constants as the ground state mesons. We therefore expect $f_{D_{s0}}
\sim f_{D_s}$ giving in addition to the heavy quark predictions 
\ber
R_{D0} \approx  R_{D1}  \approx  1 \ 
\label{oldpred}
\eer

Experimentally Belle measures \cite{belletrabelsi}
\ber 
BR[B \to D D_s(2317)]BR[D_s(2317) \to D_s \pi^0]
& = & (9.9^{+2.8}_{-2.5} \pm 3.0) \times 10^{-4} \
\label{ds}
\eer
 The dominant decay of the
$D_s(2317)$ is expected to be through the 
$D_s \pi$ mode \cite{godfreyDs,colangelo3} and so
\ber
BR[D \to D D_s(2317)] & \approx &  10^{-3}\
\label{ds1}
\eer
 Now using the measured
 branching ratio \cite{PDG}
\ber 
 BR[B^+ \to {\overline{D}}^0 D_s^+] &=& (1.3 \pm 0.4) \times 10^{-2}
\nonumber\\
BR[\overline{B}_d \to D^- D_s^+] &=& (8 \pm 3) \times 10^{-3} \
\eer
 one obtains a combined branching ratio
\ber 
 BR[B \to {D} D_s] & \approx & 10^{-2}\
\eer

This leads to $R_{D0} \approx { 1 \over 10}$ (or,$ f_{D_{s0}} \sim {1
  \over 3} f_{D_s}$) which is a factor 10 smaller then theoretical
expectations. There are a few possible explanations that can be put
forward to explain this discrepancy between experiment and theoretical
expectation and we will consider them now.

It is possible that the estimate of the decay constants of the p-wave,
$j_q= {1 \over 2}$ states in the various models are incorrect just
like the mass predictions of these states are incorrect.  This would
require a major revision of model calculations that predict the
properties of these states. From the experimental data we have seen
that $ f_{D_{s0}} \sim {1 \over 3} f_{D_s}$ which gives,
using $ f_{D_{s0}} = f_{D_{s1}^*}$, 
\ber 
R_{D1} & \approx & { 1 \over 10}\
\label{newpred}
\eer
To check this we note that experimentally Belle measures \cite{belletrabelsi}
\ber 
BR[B \to D D_s(2460)]BR[D_{s}(2460)] \to D_s \pi^0]
&=&(25.8 ^{+7.0}_{-6.0} \pm 7.7) \times 10^{-4} \nonumber\\
BR[B \to D D_{s}(2460)]BR[D_{s}(2460)] \to D_s \gamma]
& = & (5.3 ^{+1.4}_{-1.3} \pm 1.6) \times 10^{-4} \
\eer
Taking the central values we find
\ber
BR[B^+ \to {\overline D}^0 D_{s}(2460)] & \le & 31.1 \times 10^{-4}\
\label{dsstar}
\eer
 Using the measured
 branching ratio \cite{PDG}
\ber 
 BR[B^+ \to {\overline{D}}^0 D_s^{+*}] & = & (9 \pm 4) \times 10^{-3}
\nonumber\\
 BR[\overline{B}_d \to D^- D_s^{+*}] &=& (1.0 \pm 0.5) \times 10^{-2}\
\eer 
 one can obtain, using the measured central values
\ber 
 BR[B \to {D} D_s^*]  & \approx &  10^{-2}\
\label{2bodystar}
\eer
 This then leads to
$R_{D1} \approx  {1 \over 3}$ which is in disagreement with Eq. \ref{oldpred}
 and Eq. \ref{newpred}.

One might argue that factorization is not applicable to 
$ B \to D^{(*)}D^{(*)}$ decays. However recent analysis in 
Ref. \cite{luorosner} find that factorization works well for these decays.
Moreover the quantities in  Eq. \ref{ratios} are ratios of nonleptonic
decay amplitudes and so nonfactorizable effects may cancel. So what one really
requires is significantly different nonfactorizable corrections
between decays with the p-wave states in the final state and 
decays with the ground state mesons in the final state.
It is possible that the discrepancies between experiments and theory may 
arise from a combination of incorrect model prediction of p-wave 
state properties and nonfactorizable effects.

\section{Nonleptonic decays involving $B_s$ decays}

Another test of the the nature of the newly discovered $D_s$ states that 
does not rely on factorization or heavy quark symmetry involves the
$B_s$ mesons. As we indicated earlier, with the $B_s$ meson, the p-wave 
$D_s$ states can be produced via the weak current involving the $b$ quark.
We can now consider decays of the type
$B_s \to D_{s}(2317)(D_{s}(2460))M$
where $M= \pi, \rho, K $ etc. With the identification of
 $D_{s}(2317)(D_{s}(2460))$ as the p-wave states these decays are
the same as
$B_s \to D_{s0}(D_{s1}^*)M$.

One can now construct the ratios
\ber
T_{D_s(2317)}= \frac{BR[B_s \to D_{s}(2317) M]}{BR[B_d \to D_{d0}M]} \nonumber\\
T_{D_s(2460)}= \frac{BR[B_s \to D_{s}(2460) M]}
{BR[B_d \to D_{d1}^*M]} \nonumber\\
T_{D_s}= \frac{BR[B_s \to D_{s}M]}{BR[B_d \to  D_d M]} \nonumber\\
T_{D_s^*}= \frac{BR[B_s \to D_{s}^*M]}{BR[B_d \to  D_d^* M]} \
\label{sratios}
\eer
 Now in the $SU(3)$ limit all the ratios are unity. 
Moreover the ratio of ratios $r_0= { T_{D_s(2317)} \over T_{D_s}}$ and
$r_1= { T_{D_s(2460)} \over T_{D_s^*}}$  are expected to have
smaller flavour symmetry violations and hence smaller deviations from unity, 
 as $SU(3)$ breaking effects in 
the ratios may cancel \cite{datta-alpha}.
Hence any large 
deviation of $T_{D_s(2317)}$ and $T_{D_s(2460)}$ 
 from unity would be inconsistent with the $ j_q= { 1 \over 2}$ p-wave
interpretation of the new $D_s$ states. Note that the further assumption
of factorization leads to $T_{D_s(2317)} \approx T_{D_s(2460)}$
and $T_{D_s} \approx T_{D_s^*}$ in the heavy quark limit.

As indicated earlier, among various other suggestions for 
the nature of the new $D_s$ states
is the idea that these states may be $D^{(*)}K$ molecules. There are
no serious models of  such meson molecules that one can use to calculate
nonleptonic decays involving these states. Here we will attempt a rough
qualitative estimate of nonleptonic decay rates assuming that the 
$D_s(2317)$ and $D_s(2460)$ states are really a $DK$ molecule
and a $D^*K$ molecule respectively.
 Consider the nonleptonic decay $ B \to D D_s(2317)$. We assume that the 
decay proceeds 
through two stages: the first stage is the decay $B \to DDK$, 
followed by the state $D(p_2)K(p_K)$ forming the molecule $D_s(2317)$ 
with the probability 
given by
$f(p_2,p_K)$ so that 
\ber
d\Gamma( B \to D D_s(2317)) & = & 
\frac{1}{(2 \pi)^3} \frac{1}{8M_B}
|A( B \to D(p_1)D(p_2)K(p_K))|^2 f(p_2,p_K) dE_{K}dE_{2} \
\label{master}
\eer
Without a model for $f(p_2,p_K)$ we cannot make predictions
but nonetheless 
it is useful to define the average probability function
${\overline f}$ as
\ber
{\overline f} & = & \frac{\int{|A( B \to D(p_1)D(p_2)K(p_K))|^2 f(p_2,p_K) dE_{K}dE_{2}}} 
{\int{|A( B \to D(p_1)D(p_2)K(p_K))|^2  dE_{K}dE_{2}}} \
\label{fbar}
\eer
Hence we have
\ber
BR( B^+ \to \overline{D}^{(0*)} D_{s}(2317)^+) & = &
 BR( B^+ \to \overline{D}^{(0*)} D^{+} K^0) \times {\overline f} \nonumber\\
BR( B^0 \to {D}^{(-*)} D_{s}(2317)^+) & = &
 BR( B^0 \to {D}^{(-*)} D^{0} K^+) \times {\overline f} \
\eer
We can define a similar function $f^*$ and the average $ {\overline f}^*$
for nonleptonic decays involving the $D_s(2460)$ and so
\ber
BR( B^+ \to \overline{D}^{(0*)} D_{s}(2460)^+) & = &
 BR( B^+ \to \overline{D}^{(0*)} D^{+*} K^0) \times {\overline f}^* \nonumber\\
BR( B^0 \to {D}^{(-*)} D_{s}(2460)^+) & = &
 BR( B^0 \to {D}^{(-*)} D^{0*} K^+) \times {\overline f}^*\
\label{fbarstar} 
\eer
We can consider the ratios
\ber 
Z_{res}^{+} =  \frac{BR( B^+ \to \overline{D}^{0} D_{s}(2460)^+)} 
{BR( B^+ \to \overline{D}^{0} D_{s}(2317)^+)} ,Z_{res}^{+*} =  \frac{BR( B^+ \to \overline{D}^{0*} D_{s}(2460)^+)} 
{BR( B^+ \to \overline{D}^{0*} D_{s}(2317)^+)} \nonumber\\
Z_{res}^{0}  =  \frac{BR( B^0 \to {D}^{-} D_{s}(2460)^+)} 
{BR( B^0 \to \overline{D}^{-} D_{s}(2317)^+)},Z_{res}^{0*}  =  \frac{BR( B^0 \to {D}^{-*} D_{s}(2460)^+)} 
{BR( B^0 \to \overline{D}^{-*} D_{s}(2317)^+)} \nonumber\\
Z_{3-body}^{+} =  \frac{BR( B^+ \to \overline{D}^{0} D^{+*} K^0)} 
{BR( B^+ \to \overline{D}^{0} D^{+} K^0)}, 
Z_{3-body}^{+*}  =  \frac{BR( B^+ \to \overline{D}^{0*} D^{+*} K^0)} 
{BR( B^+ \to \overline{D}^{0*} D^{+} K^0)} \nonumber\\ 
Z_{3-body}^{0} = \frac{BR( B^0 \to {D}^{-} D^{0*} K^+)} 
{BR( B^0 \to \overline{D}^{-} D^{0} K^+)}\,
Z_{3-body}^{0*} =  \frac{BR( B^0 \to {D}^{-*} D^{0*} K^+)} 
{BR( B^0 \to \overline{D}^{-*} D^{0} K^+)}\
\label{res-threebody}
\eer
which are related as
\ber
Z_{res}^{+}= Z_{3-body}^{+}\frac{\overline{f}^*}{\overline{f}}, Z_{res}^{+*}= Z_{3-body}^{+*}\frac{\overline{f}^*}{\overline{f}}\nonumber\\
Z_{res}^{0}= Z_{3-body}^{0}\frac{\overline{f}^*}{\overline{f}}, Z_{res}^{0*}= Z_{3-body}^{0*}\frac{\overline{f}^*}{\overline{f}}\
\eer
Using the measured three body branching ratios \cite{babarDDK}
\ber
BR( B^+ \to \overline{D}^0 D^{+} K^0) &=&
(0.18 \pm 0.07 \pm 0.04) \times 10^{-2}\nonumber\\
BR( B^0 \to {D}^{-} D^{0} K^+) &=&
(0.17 \pm 0.03 \pm 0.03) \times 10^{-2}\nonumber\\
BR( B^+ \to \overline{D}^{0*} D^{+} K^0) & = &
(0.41^{+0.15}_{-0.14} \pm 0.08 ) \times 10^{-2} \nonumber\\ 
BR( B^0 \to {D}^{-*} D^{0} K^+) & = &
(0.31^{+0.04}_{-0.03} \pm 0.04 ) \times 10^{-2} \nonumber\\ 
BR( B^+ \to \overline{D}^0 D^{+*} K^0) & = &
(0.52^{+0.10}_{-0.09} \pm 0.07 ) \times 10^{-2} \nonumber\\
BR( B^0 \to {D}^- D^{*0} K^+) & = &
(0.46 \pm 0.07 \pm 0.07 ) \times 10^{-2} \nonumber\\
BR( B^+ \to \overline{D}^{0*} D^{+*} K^0) & = &
(0.78^{+0.26}_{-0.21} \pm 0.14 ) \times 10^{-2}\nonumber\\
BR( B^0 \to {D}^{-*} D^{0*} K^+) & = &
(1.18 \pm 0.10 \pm 0.17 ) \times 10^{-2}\
\label{3body}
\eer
which are proportional to either $1 - \overline f$ or $1 - {\overline f}^*$
and
 assuming $ {\overline f}^* \approx  {\overline f}^*$ allows one 
to obtain, with the central values of the measurements,
\ber
Z_{3-body}^+ & = & 2.89
\eer
which can be compared to  $Z_{res}^+ =3.14$ from Eq. \ref{ds1} and 
Eq. \ref{dsstar}. If fact the prediction
$Z_{res}^{+} \sim 3$ , $Z_{res}^{+*} \sim 3$, 
 $Z_{res}^{0} \sim 3$ and $Z_{res}^{0*} \sim 3$  
are consistent within
 the errors for the three body branching ratios in
$Z_{3-body}^{+} $, $Z_{3-body}^{+*} $, 
 $Z_{3-body}^{0} $ and $Z_{3-body}^{0*} $.
 We also obtain
 $ {\overline f} \approx  {\overline f}^* \approx $ 0.3 , from
 Eq. \ref{ds1} and Eq. \ref{3body} which indicates
that a sizeable fraction of the $D^{(*)}K$ state form molecules.
 
Finally we can extend this model also to the case where the $D_s$ resonance 
is produced via the weak current containing the $b$ quark in $B_s$ decays.
Consider the decays $B_s \to D_s(2317) M$ where $M$ is the emitted meson.
The form factor for $ B \to D_s(2317)$ transition can then be related to
$B \to D K $ transition. In other words, we can write
\ber
BR[B_s \to D_s(2317) M] & = & BR[B_s \to D K M] \overline{f}^{\prime}\
\eer  
where 
\ber
\overline{f}^{\prime} & = & \frac{\int{|A( B \to D(p_2)K(p_K)M(p_1))|^2 f(p_2,p_K) dE_{K}dE_{2}}} 
{\int{|A( B \to D(p_2)K(p_K)M(p_1))|^2  dE_{K}dE_{2}}} \
\label{fbar}
\eer
We can similarly define 
$\overline{f}^{*\prime}$ as
\ber
BR[B_s \to D_s(2460) M] & = & BR[B_s \to D^* K M] \overline{f}^{*\prime}\
\eer    
Note that the ratios $T_{D_s(2317)}$ and $T_{D_s(2460)}$ (Eq. \ref{sratios}) in the molecular model 
are no longer equal to unity in the $SU(3)$ limit since that depended
on the identification of these states as p-wave states. Therefore the measurement of these ratios can provide useful information on the nature of the
$D_s(2317)$ and the $D_s(2460)$ states.

\section{Summary and Conclusions}

In summary, in this work, we have considered the nonleptonic $B$ decays
$ B \to D^{(*)} D_s(2317)(D_s(2460))$,
 involving the newly discovered 
$D_s(2317)$ and the $D_s(2460)$ states. We have discussed the implication of
the measured nonleptonic decays for the properties and the nature of 
these states. If these 
states are the  p-wave multiplet with the light degrees of freedom in
the $ j_q = { 1 \over 2}$ state, then we find that experiments indicate
disagreement with model calculation of their properties and/or breakdown of the factorization assumption. 
 We have suggested  further tests
involving nonleptonic $B_s$ meson decays,
 that do not assume factorization but assumes
 $SU(3)$ flavour symmetry,
 that can further shed light on the true nature of these 
newly discovered states. Finally,
we have also proposed a model
to calculate the
two body nonleptonic decays $ B \to D^{(*)} D_s(2317)(D_s(2460))$,
assuming that the the $D_s(2317)$ and $D_s(2460)$ are 
$DK$ and $D^*K$ molecules. The model relates these two body nonleptonic decays
 to the three body $B$ decays of the type $ B \to D^{(*)}D^{(*)}K$.

\section{Acknowledgments}

We thank Tom Browder 
for discussions. This work is supported by the Natural Science and
Engineering Council of Canada (NSERC) under grant number A3828.

\end{document}